# Thermoelectric properties of 3D topological insulator: Direct observation of topological surface and its gap opened states

Stephane Yu Matsushita,[1*] Khuong Kim Huynh,[2] Harukazu Yoshino,[3] Ngoc Han Tu,[1] Yoichi Tanabe,[1] and Katsumi Tanigaki[1,2 **]

[1]Department of Physics, Graduate School of science, Tohoku University, Sendai 980-8578, Japan

[2]WPI-Advanced Institute for Materials Research, 2-1-1 Katahira, Aoba-ku, Sendai, Miyagi, 980-8578, Japan

[3]Division of Molecular Materials Science, Graduate School of Science, Osaka City University, Osaka 558-8585, Japan

*E-mail address: *m.stephane@m.tohoku.ac.jp, **tanigaki@m.tohoku.ac.jp

**Abstract**

We report thermoelectric (TE) properties of topological surface Dirac states (TSDS) in three-dimensional topological insulators (3D-TIs) purely isolated from the bulk by employing single crystal $Bi_{2-x}Sb_xTe_{3-y}Se_y$ films epitaxially grown in the ultrathin limit. Two intrinsic nontrivial topological surface states, a metallic TSDS (m-TSDS) and a gap-opened semiconducting topological state (g-TSDS), are successfully observed by electrical transport, and important TE parameters (electrical conductivity ($\sigma$), thermal conductivity ($\kappa$), and thermopower ($S$)) are accurately determined. Pure m-TSDS gives S=-44 $\mu VK^{-1}$, which is an order of magnitude higher than those of the conventional metals and the value is enhanced to -212 $\mu VK^{-1}$ for g-TSDS. It is clearly shown that the semi-classical Boltzmann transport equation (SBTE) in the framework of constant relaxation time ($\tau$) most frequently used for conventional analysis cannot be valid in 3D-TIs and strong energy dependent relaxation time $\tau(E)$ beyond the Born approximation is essential for making intrinsic interpretations. Although σ is protected on the m-TSDS, $\kappa$ is greatly influenced by the disorder on the topological surface, giving a dissimilar effect between topologically protected electronic conduction and phonon transport.

Realization of high performance thermoelectric (TE) materials[1] is an important challenge in science and technology and a frontier research area for solving the future energy problems. Performance of TE materials can generally be characterized by the dimension-less figure of merit ($ZT = S^2(\sigma/\kappa)T$), where $\sigma$ is the electric conductivity, $\kappa$ the thermal conductivity and $S$ the thermoelectric power (so called Seebeck coefficient). In order to improve the performance of TE materials, it is consequently necessary to increase the thermoelectric power factor $P=\sigma S^2$, which is mainly associated with the electronic part of materials. At the same time, the ratio of $\sigma/\kappa$, where $\kappa=\kappa_e+\kappa_{ph}$ consists of both electron (e) and phonon (ph) contribution, becomes important simultaneously from the viewpoint of thermodynamic efficiency. Because the term of $(\sigma/\kappa)T=1/L$ is recognized to be constant in many conventional materials as far as the normal electronic states are concerned, being known as the Wiedemann-Franz Law (WFL) with $L$ as the Lorenz number[2], a way of research for breaking down the WFL is also of significance for exploring high performance TE materials.

The improvement in efficiency of TE materials has intensively been explored in these decades with the strategy of designing low-$\kappa$ materials under the concept of phonon-glass-electron-crystal (PGEC)[3-6]. The dimensionality control and the quantum size effect in nanostructure have also been other important approaches for achieving higher $ZT$[7,8]. The recent discovery of topological insulators (TIs) provides a new research platform in this research field. TIs are attracting considerable attention in contemporary materials science showing gapless helical massless Dirac fermions on a two-dimensional (2D) surface or one-dimensional (1D) edge[9-11]. Although there is no direct relationship, TIs and good TE materials have common features of their properties and actually typical three dimensional TIs (3D-TIs) such as $Bi_2Se_3$, $Bi_2Te_3$ and $Bi_{2-x}Sb_xTe_3$ had been studied as a promising candidate for TE materials in the past[12-18].

Recent theoretical studies suggested that additional nontrivial conduction channels existing in pure topological surface Dirac states (TSDS) in three-dimensional TIs (3D-TIs) may provide a unique route to enhance $ZT$. Numerous theoretical works debate this intriguing scientific issue on the nontrivial metallic TSDSs (m-TSDS)[19-25]. Two possible mechanisms are proposed for enhancing the $ZT$ in TIs: (i) Strong energy dependence of relaxation times in m-TSDS can lead to large and anomalous Seebeck effect via electron and hole asymmetric contribution; (ii) Intermediate gap-opened surface states (g-TSDS) at the neutral Dirac point (NDP) via hybridization between the top and the bottom topological surface with partly leaving spin momentum locking texture before full gap opening could enhance the $ZT$[26,27]. Especially, the former theoretical prediction is greatly different from the common understanding of TE properties, because $S$ (consequently $ZT$) of metallic states is in principle very small in conventional metals and cannot be considered as a good target for TE materials. Although many experimental approaches to electrical transport measurements have been reported so far[28-30], no direct and firm observations on the TE properties of 3D-TIs for pure m-TSDS as well as g-TSDS have successfully been provided due to the difficulty in the separation of the topological surface state from the bulk.

Here, we report the intrinsic TE properties of 3D-TIs by targeting on high quality tetradymite $Bi_{2-x}Sb_xTe_{3-y}Se_y$ (BSTS) single crystal thin films. In order to study the intrinsic TE parameters, we grow 3D-TI BSTS thin films with thickness ranging from the ultrathin limit of 4 QLs to a nearly bulk of 40 QLs. We successfully observe purely isolated m-TSDS at 8QL by decreasing the contribution of the bulk carriers in the ultrathin film limit. We also find g-TSDS in the limit of 4QL. Importantly, the grown single crystal BSTS thin films can beneficially be transferred to any other substrates under a damage-free condition[31,32], and therefore physical parameters important for having intrinsic interpretations on the TE properties can be available by various measurements. The value of $S$ for m-TSDS is found to be one order larger than that of conventional metals. We clearly show that any theoretical models proposed so far cannot explain the TE parameters experimentally determined in our present studies, and suggest that strongly energy-dependent relaxation times ($\tau(E)$) beyond the Born approximation should be taken into account for real interpretations.

BSTS single crystal thin films with 1cm$^2$ large in size were grown on mica substrate with a catalyst-free epitaxial physical vapor deposition (PVD) method using a dual-quartz system, which detail method was reported elsewhere[31,32]. A highly insulating $Bi_{1.5}Sd_{0.5}Te_{1.7}Se_{1.3}$ single crystal was used as a source material[33,34], and the film thickness was controlled by the deposition time. The quality of the grown films was characterized by energy dispersive X-ray (EDX) spectroscopy, Raman spectroscopy and X-Ray diffraction (XRD). The thickness of the film was measured by atomic-force-microscopy (AFM). The TE transport was measured by using a home-built device by a steady-state method. The analytical details were described elsewhere[35]. All measurements were carried out by using Physical Properties Measurement System (PPMS, Quantum Design) under a high-vacuum condition to minimize the thermal conduction by the remaining gas.

The structure of BSTS having periodical five atomic-layers (1QL=0.78 nm) and its schematic band diagram are shown in Fig. 1(a) and (b). The detailed band diagram required for interpretations of the TE properties was determined based on the quantum oscillations as well as the evolution of resistivity ($\rho$) as a function of temperature ($T$) in addition to the previous literatures as described later[31,32,34]. As schematically illustrated in Fig. 1(b), the chemical potential $\mu$ resides 59-85 meV above the Dirac neutral point (DNP) in the n-type m-TSDS inside the energy gap of the BSTS bulk bands (the band gap $\Delta E_B(=E_c-E_v)=0.3$ eV[34]). The p-type impurity level $E_i$ exists above the valence band maximum (VBM) in the energy scale of 3-6 meV, and therefore itinerant holes are thermally populated in the VB at high-$T$ in a thick sample. This influence of bulk impurity level can greatly be minimized by reducing the film thickness, and the intrinsic metallic electronic states of m-TSDS were observed for 8QL-BSTS. When the film thickness became further thinner to 4QL, an energy gap opening on the m-TSDS, being caused by the hybridization of wave functions between the top and the bottom m-TSDS, was observed.

Figure 2(a) shows $T$ evolution of the 2D sheet resistances ($R_\square$) of four BSTS thin films with selected important thicknesses (40QL, 30QL, 8QL, and 4QL). It is important to see that $R_\square$ at 300 K increased with a decrement in film thickness $t$: $R_\square$=5.5 (40 QL), 11.0 (30 QL), 22.2 (8 QL), and 104 k$\Omega$ (4 QL). The values of bulk conductance $G$ observed for BSTS with various film thicknesses were analyzed by employing a two-layer parallel connection circuit model described as: $G = G_\square^s + G_\square^b t$ (1), where $G_\square^s$ and $G_\square^b$ are the sheet conductance of the topological surface and the bulk, respectively. A typical insulating property of bulk 3D-TIs was observed for 40QL-BSTS, where semiconducting $R_\square$ reached a maximum at around 50 K and started to decrease again as $T$ became further low. Intriguingly, by reducing the film thickness, a metallic $T$ dependence arising from the intrinsic nontrivial m-TSDS was successfully observed for 8QL-BSTS. The confirmation of pure m-TSDS is also given later by the $T$-linear dependence of $S$. When the film thickness is further decreased to 4QL, its $R_\square$ intriguingly showed to increase again at low-$T$, being indicative of an energy gap opening in m-TSDS. The activation energies ($\Delta Ea$) were estimated from the Arrhenius plots of $R$ to be 6.9 meV, 3.2 meV and 0.15 meV for 40QL, 30QL and 4QL films, respectively. The quite small value of $\Delta E_a$ for 4QL-BSTS can be attributed to the surface gap state (g-TSDS) created via hybridization of the wave functions of the two m-TSDSs[36].

Figure 2(b) shows $S$ values of these four BSTS thin films. 40QL-BSTS showed p-type $S$ with a nonlinear $T$ dependence in the plus charge polarity, indicating that the carriers are thermally populated holes from the impurity level $E_i$, followed by saturation with a maximum value of 170 μVK$^{-1}$ at 300 K. On the other hand, 8QL-BSTS (m-TSDS) showed $S$ of -44 μVK$^{-1}$ with minus charge of electrons at 300 K and its $T$ dependence is in a linear fashion, being indicative of the n-type intrinsic metallic electronic states of m-TSDS as described earlier. $S$ values of conventional metals are generally on the order of 1 μVK$^{-1}$, and intriguingly the observed value is an order larger than those of conventional metals[37]. This unusually large $S$ value in m-TSDS is most likely associated with the unconventional linear Dirac dispersion surface band having one to one correspondence with the bulk, and the detailed discussion will be given later. It can also be seen that $S$ of 4QL-BSTS (g-TSDS) greatly increased to a larger value of -212 μVK$^{-1}$ with the minus charge polarity in a nonlinear fashion. This gives the unambiguous experimental observation detected in electrical transport that m-TSDS of 8QL-BSTS becomes gap opened in the case of 4QL due to the hybridization of wave functions between the top and the bottom m-TSDS, as suggested by ARPES experiments[38,39]. In the case of thick 30QL-BSTS, $S$ showed a mixed state of p-type bulk state

(BS) and n-type m-TSDS. This can further be supported by the experimental fact that the $T$ gradient of $S$ is the same between 30QL- and 8QL-BSTS in the low-$T$ limit.

$T$ dependence of $\kappa$ for 40QL (nearly bulk), 8QL (m-TSDS) and a bulk single crystal BSTS (200 μm in thickness) is shown in Fig. 3. The $\kappa$ values of BSTS thin films were obtained by subtracting the value of $\kappa$ of a mica substrate from the total value of $\kappa$. This correction was essential because thermal energy flows in both BSTS thin films and a mica substrate, while the situation is markedly different from that of electrical current only flowing in metallic substances. As one can see in Fig.3, the $\kappa$ values of 40QL-BSTS and bulk BSTS were almost the same with each other, indicating that the thermal transport of bulk carriers is still dominant in a sample as thin as 40QL, the situation of which was similar to the electrical transport experiments described earlier. The value of $\kappa$ for 40QL-BSTS at 300 K was 1.7 W K$^{-1}$ m$^{-1}$ and close to those of other reports in literature[40]. The $\kappa$ of 1.5 W K$^{-1}$ m$^{-1}$ at 300 K for m-TSDS of 8QL-BSTS was comparable with or a little bit smaller than those of the bulk single crystal and 40QL-BSTS.

According to the highly insulating bulk state of BSTS, the contribution of phonons is dominant for the $\kappa$ of bulk and 40QL-BSTS. Additionally, if the electronic thermal conductivity $\kappa_e$ is dominant, the temperature dependence of $\kappa$ should be $T^2$, but this was not the case in the present experiments. Consequently, the comparable value of $\kappa$ of 8QL-BSTS to the bulk one indicates that phonons contribute dominantly to $\kappa$ for m-TSDS. Importantly to be noted here is the fact that $\kappa$ of 8QL decreases much greater at low-$T$ than those of the bulk and 40QL-BSTS without showing any specific peaks around 20 K. In general, a maximum peak in $\kappa$ observed at low $T$ is an important typical phenomenon to be observed for single crystals and can be attributed to the crossover of the competing contributions between the phonon mean free path becoming longer and the loss in the number of the phonon modes participating in $\kappa_{ph}$ as $T$ decreases[37]. The loss of such a maximum peak in $\kappa$ at around 20 K observed for m-TSDS is an important experimental evidence of phonon anharmonicity, indicating that phonons are greatly scattered on the disordered surface of 3D-TIs[41-43]. Although the transport of Dirac electrons is topologically well protected in m-TSDS, the phonon transport is not protected in a similar fashion on its disordered surface, which is important experimental evidence for the breakdown in Wiedemann-Franze law (WFL).

In order to discuss quantitatively the $S$ values (absolute values and sign) experimentally determined, the accurate band picture of BSTS thin films described in Fig. 1(b) was evaluated from the physical parameters as described below and the band gap of BSTS determined by ARPES[34].

The p-type impurity level was determined to be 6.9 meV above the valence band maximum (VBM) for 40QL-BSTS, and 3.2 meV for 30QL-BSTS from the $T$ dependence of $R$. The chemical potential of m-TSDS can be estimated by SdH oscillations. The 2D carrier density ($n_{2D}$) was evaluated to be 1.68 × 10$^{12}$ cm$^{-2}$ (69 T) for 30QL-BSTS, and 7.74 × 10$^{11}$ cm$^{-2}$ (32 T) for 8QL-BSTS, respectively. Using the value of effective cyclotron mass $m_c = 0.124\ m_e$, the chemical potential $\mu=\hbar 2\pi n_{2D}/m_c$, was evaluated to be 129 meV for 30QL-BSTS, and 59 meV for 8QL-BSTS locating above the Dirac point (DP), respectively. Applying the value of $v_F = 4.0\times10^5$ m s$^{-1}$ reported by ARPES[34], we can also obtain the $\mu$ value to be 120 meV and 81 meV for 30QL and 8QL-BSTS, respectively. These values evaluated by the two different experimental methods for BSTS samples made by different two research groups were almost identical. The $\mu$ of m-TSDS importantly lies below the middle in the bulk band gap and above the DNP, where the m-TSDS band is occupied by electrons.

The observed TE properties of pure m-TSDS allows us to discuss the accuracy of theoretical predictions. TE properties of 3D-TIs are mainly discussed in the framework of a semi-classical Boltzmann transport equation (SBTE)[19-25]. For the conventional materials, the physical parameters required for comparing the experimental data to the theoretical calculations are the chemical potential $\mu$ and relaxation time $\tau$. In the case of 3D-TIs, since two types of conduction channels of surface and bulk exist, the ratio of surface-to-bulk can be considered to affect sensitively to $\sigma$, $S$, and $\kappa$. Therefore, in order to discuss the applicability of theoretical models and to understand the intrinsic nature of TE properties of 3D-TIs, it is important to discuss these points by focusing on the pure m-TSDS.

Here, we compare our result of m-TSDS with the three types of theoretical models so far proposed, which treat the relaxation time in a different manner. First one was proposed by Tretiakov *et al.*[22], where a constant relaxation time ($\tau_0$) is used, and S is written as:

$$S = -\frac{1}{eT}\frac{L_1}{L_0} \quad (3)$$

$$L_j = \frac{1}{8\pi^2\hbar^2}(k_BT)^{j+1}\int_{\overline{\Delta}_s}^{\infty}\tau x\left[\frac{(x-\bar{\mu})^j}{\cosh\frac{x-\bar{\mu}}{2}} + (-1)^j\frac{(x+\bar{\mu})^j}{\cosh\frac{x+\bar{\mu}}{2}}\right]dx \quad (4),$$

where $x = E/k_BT$, and $\bar{\mu} = \mu/k_BT$. The second is a two-$\tau$ model proposed by Xu et al.[21], where two different relaxation times of $\tau_1$ and $\tau_2$ are considered, one corresponding to inside the bulk band gap ($\tau_1$) and the other being above (below) the conduction (valence) band ($\tau_2$). In the first model, the only experimental parameter for calculation is $\mu$, while in the latter one, an additional fitting parameter $r=\tau_1/\tau_2$ (the ratio of the two relaxation times) is needed. We employed the experimental value of $\mu = 81$ meV estimated from SdH measurements as described earlier. The results are shown in Fig. 4(a). The constant $\tau$ model shows an almost linear T-dependence, but a large discrepancy is evident when it is compare to the experimental values. In the case of the two-$\tau$ model, $S$ decreases at high-$T$ and becomes non-linear as a function of $T$, being not consistent with the present experimental data. Both of them can not explain the observed S and this provides an important message that a more sophisticated models must be applied for understanding the thermoelectric properties of 3D-TI BSTS. Another way of handling the relaxation time was proposed in the framework of the first order Born approximation $\tau \propto E^{-1}$ by Takahashi and

Murakami[23]. In this model, $S$ shows an opposite sign between the surface and the bulk states in the vicinity of the band even if the carriers are the same in type because of the significantly different relaxation times of the two states. However, this contradicts our present experimental observations. As a modified relaxation time in a realistic fashion, we preliminarily supposed a more general energy-dependent relaxation time approximation described by $\tau = \frac{1}{c \cdot (E/k_B T)^r}$, where c and r are the constants independent of $T$ and $E$[44,45]. A reasonable fit was obtained for c=1 and r=0.5 under the intrinsic experimental parameters of $\mu$= 81 meV as can be seen in Fig. 4(c) as the red line.

The discussion made so far indicates that the energy dependent relaxation times beyond the first Born approximation is necessary. Considering that TSDS can be created via one to one topological correspondence between the bulk and the surface, it is very reasonable that the relaxation of itinerant electrons is strongly dependent on the chemical potential $\mu$. These results can give a useful guideline for designing a new theoretical model to understand the TE properties of 3D-TIs.

Let us now discuss the ZT values of TSDS. Given $\sigma = G_\square^s/t_s$, where $G_\square^s = 4.2 \times 10^{-5}$ $\Omega^{-1}$ and $t_s = 0.16$ nm for one atomic layer in 1QL of BSTS using the TE parameters of $\sigma = 2.7 \times 10^5$ S m$^{-1}$, $\kappa$=1.5 W m$^{-1}$ K$^{-1}$, and $S$=-44 µV K$^{-1}$ experimentally determined for the intrinsic m-TSDS of 8QL-BSTS in the present work, $ZT$ can be evaluated to be 0.1 at 300 K. The figure of merit $ZT$ for various metallic materials is depicted in Fig. 4(b). Generally in normal metals such as Ni and Pt, $ZT$ values are very small on the order of 0.01 due to the small $S$. It is very intriguing to know that m-TSDS with Dirac Fermions can provide a large $ZT$ even though it is purely metallic.

A similar enhancement of ZT in metallic systems has been found for Co Oxides[46-48] and Weyl/Dirac semimetals[49]. The former is categorized as a strong correlated electron system. The latter has a similar linear dispersion band in the bulk to that on the topological surface. However, it should be noted that the physics of 3D-TIs and Weyl/Dirac semimetals is not the same with each other. There is no classification of 2D Dirac/Weyl semimetals in the general classification table[50,51], while Dirac electrons of 3D-TI are confined in the 2D-surfac with the one-to-one correspondence to the bulk. We compare the ZT value of m-TSDS with those of other metallic thermoelectric materials in Fig. 4(b). Intriguingly, Dirac/Weyl metallic bands and 3D-TIs seem to give the highest ZT values among these families although they are purely in the metallic regime. In order to realize high performance of TE materials, the factors giving the special electronics states breaking down the conventional metallic states are required, such as strong electron correlations for $SrTiO_3$[52] and orbital freedom for $CoO_2$. The high mobility as well as the unique energy dependent relaxation times beyond the Born approximation in the topologically protected m-TSDS of 3D-TIs can be a new route to enhance the ZT value in a metallic system.

In addition, we pointed out earlier that $\kappa$ is greatly affected from the disorder on the TSDS. The Lorentz number of 8QL-BSTS can be estimated to be L = $1.7 \times 10^{-8}$ W$\Omega$K$^{-2}$ at 300 K, which shows a 30% reduction from the general value ($2.44 \times 10^{-8}$ W$\Omega$K$^{-2}$). Although the electronic states of m-TSDS are topologically protected, the phonon transport is not protected in a similar fashion on the disordered surface. It is noted the highest thermoelectrics performance of Bi$_2$Te$_3$ and its family of compounds is only realized in bulk polycrystals, but not in single crystals, by carefully controlling grain boundaries using spark plasma sintering in the engineering manner[14,53]. It is intriguing to see in the future whether a high ZT value can be seen in single crystal 3D-TIs when the complex spin texture is controlled before the full gap opening in the intermediate gap opening states of g-STDS.

We successfully observed TE parameters of purely isolated m-TSDS and g-TSDS in 3D-TI BSTS by employing high quality ultra-thin films in the extreme limit of 4 to 8 QL with 1cm$^2$ in size grown by a vapor phase epitaxial method. In BSTS having a different carrier type between m-TSDS having pure 2D electronic states and the 3D bulk, we directly determined the intrinsic TE parameters of m-TSDS for 8QL-BSTS. While m-TSDS is purely metallic with single Dirac band on the surface, the thermopower *S* and therefore *ZT* is an order of magnitude larger than those of conventional metals. This can be interpreted in terms of the strong energy dependent relaxation times ($\tau(E)$) arising from the bulk-surface 1:1 topological correspondence in 3D-TIs.-The observed TE parameters were discussed in comparison with the proposed theoretical models in the framework of SBTE, suggesting that more sophisticated theoretical models including accurate energy dependent relaxation times beyond the Born approximation will be necessary.

**Acknowledgments**

This work was supported in part by a Grant-in-Aid for Scientific Research from the Ministry of Education, Culture, Sports, Science and Technology (MEXT). S.Y.M. thanks Tohoku University Interdepartmental Doctoral Degree Program for Multi-dimensional Materials Science Leaders for financial support.

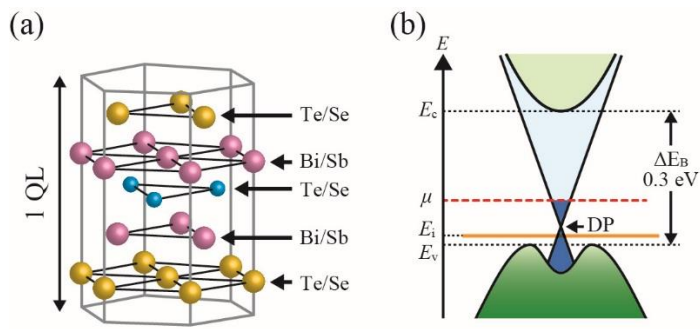

**Figure 1. Crystal structure and schematic electric band diagram of BSTS thin film.** (a) Crystal structure of BSTS. One quintuple layer (QL) has three sets of layers: Te/Se (orange and blue) and Bi/Se (pink). (b) Schematic electric band diagram of metallic BSTS (m-TSDS). The orange line shows the position of an impurity level ($E_i$), and the red dotted line indicates the position of the chemical potential ($\mu$). $\Delta E_B$ is the energy gap of the bulk state.

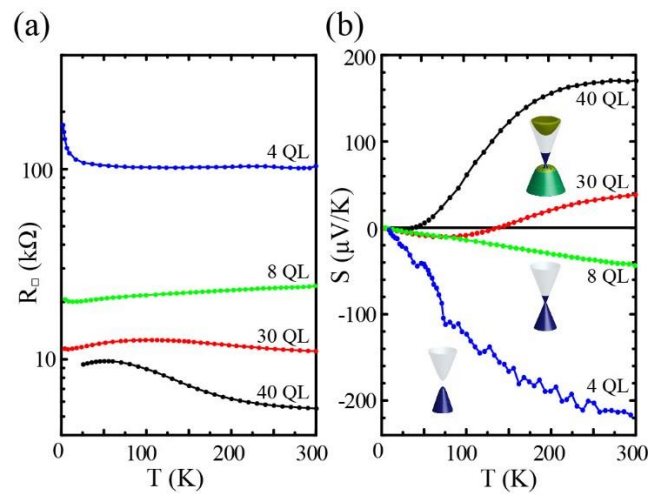

**Figure 2. Electrical transport properties of BSTS thin films.** (a) Temperature dependence of sheet resistance ($R_\square$) and (b) temperature dependence of Seebeck coefficient ($S$) of BSTS thin films of 40QL, 30QL, 8QL and 4QL. Schematic band picture of each film are shown in (b): 40QL and 30QL-BSTS are in a mixed state of a bulk state (i-BS) and a metallic topological surface Dirac state (m-TSDS), 8QL-BSTS is gapless m-TSDS, and 4QL-BSTS is gap opened TSDS (g-TSDS).

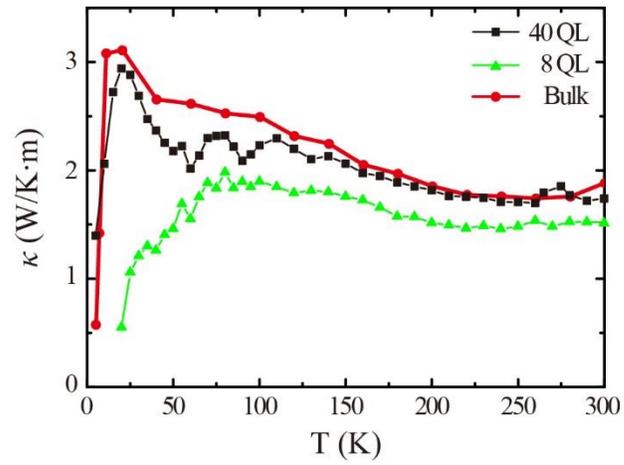

**Figure 3. Thermal conductivity of 40QL (black square) and 8QL (green triangle) BSTS thin films.** The red dots and their lines indicate T dependence of $\kappa$ for a bulk single crystal of 200 μm in thickness. A typical error bar is shown for the data of 40 QL thin film (black dots).

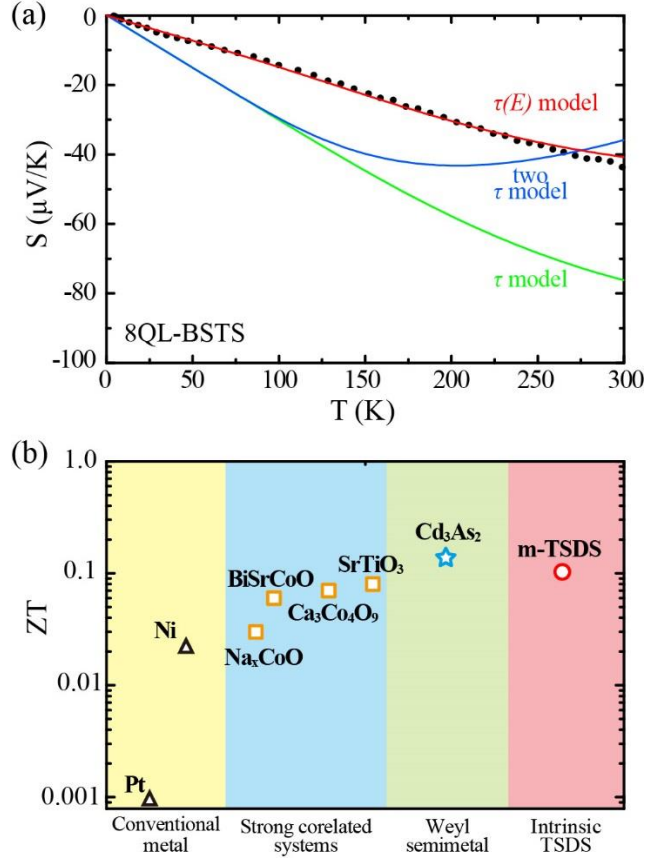

**Figure 4. Thermoelectric property of m-TSDS.** (a) Theoretical calculations of $S$ for m-TSDS of 8QL-BSTS. The red line is the best fitting of the experimental data (black dots) using an energy-dependent relaxation time ($\tau(E)$) model. The green and blue lines are simulations for the constant $\tau$ and the two-$\tau$ model, respectively, with the chemical potential $\mu$ estimated from the SdH oscillations. The details of calculations are described in the text. (b) Thermoelectric figure of merit (ZT) of various metallic systems at 300 K. The four colored areas in yellow, blue, green and red represent the $ZT$ values of conventional metals (Pt and Ni), strongly correlated electron systems ($Na_xCoO$[46], $BiSrCoO$[47], $Ca_3Co_4O_9$[48], $SrTiO_3$[52]), Weyl semimetal ($Cd_3As_2$[49]), and the intrinsic m-TSDS (present study), respectively.